\documentclass{appolb}
\usepackage{epsfig}

\begin{document}
\title{ CHARM FRAGMENTATION AND DIJET ANGULAR DISTRIBUTIONS %
\thanks{Presented on behalf of
      the ZEUS Collaboration, at DIS02 conference in Cracow, Poland.}
}
\author{Sanjay Padhi 
\address{Department of Physics, McGill University, Montreal, \\ 
Quebec, Canada, H3A 2T8 \\
 E-mail: spadhi@mail.desy.de}
}
\maketitle
\vspace{-0.8cm}
\begin{abstract}
Charm fragmentation and dijet angular distributions  have been
measured in $D^*$ photoproduction at HERA. Charm fragmentation and 
its property of universality is evaluated in terms of measurement 
of $P_v$, the ratio of vector/(vector + pseudoscalar) mesons. Angular
distributions of dijets, with at least one of the jets associated 
with a $D^{*\pm}$ meson, have been measured for
samples enriched in direct or resolved photon events. The
differential cross section shows a steep rise for 
resolved events in the photon direction, providing strong 
evidence that the bulk of the
resolved photon cross section is due to the charm content of the photon. 
The shallower
rise for direct events as well as for resolved photon
events in the proton direction are consistent with the quark exchange
diagrams.
\end{abstract}
\section{Introduction}
Heavy quark photoproduction and fragmentation at HERA offers a novel
way of testing both perturbative and non-perturbative aspects of
quantum chromodynamics (QCD). In this paper parton
dynamics (in terms of angular distributions) and fragmentation will be
addressed based on recent ZEUS
measurements.

\section{Charm Fragmentation}

 A non-perturbative aspect of QCD can be measured by considering the
 spin dependence in fragmentation, which should in principle be sensitive to
 non-perturbative effects in the hadronisation process.
ZEUS recently 
measured the fragmentation ratio
 P${_v}$, ratio of vector/(vector + pseudoscalar). The value P$_{v}$ 
can be calculated with respect to the ground-state charm mesons via the decay
channels D$^{*+}$
$\rightarrow$ D$^{0}\pi^{+}_{s} \rightarrow (K^{-}\pi^{+})\pi^{+}_{s}(+c.c.)$
and D$^{0}\rightarrow K^{-}\pi^{+}(+c.c.) $.

 In this paper, the analysis~\cite{1} is based on $D^{*\pm}$ and
 $D^{0}$ events
 with an almost real photon (virtuality, $Q^2_{median}
\sim 3.10^{-4} $ GeV$^2$) in a photon-proton center-of-mass energy,
W, in
the range $130 <$ W $< 295$ GeV. Using $\triangle$M = M(D*) -
M(D$^0$) as a $D^*$ tag, the
sample is then further divided into $D^{0}$ mesons arising from and
not from $D^*$
mesons. After this division there were $ 1180 \pm 39$ events with a $D^{0}$
meson from a $D^*$ and $ 5223 \pm 185$ inclusive $D^{0}$ meson events and
the resulting value for P$_{v}$ in the full phase space is:
\begin{center}
 P$_{v}$ = 0.546 $\pm 0.045(stat)^{+0.028}_{-0.028}(syst)$.
\end{center}
This is in good agreement with the values of $ 0.57 \pm 0.05 $
and $0.595 \pm  0.045 $~\cite{2} measured in e$^+$e$^-$ annihilation.

\section{Dijet in charm photoproduction}
The angular distribution of outgoing partons in a hard partonic
process is an efficient tool to study the parton dynamics of the
underlying sub-processes. In leading order (LO) QCD these underlying 
sub-processes (Fig.1) can be divided into either direct photon or  
resolved photon processes. In direct photon processes the photon
participates in the hard scatter predominantly via the boson-gluon
fusion process. This process has a quark as the propagator in the hard
interaction. In resolved photon
processes the photon acts like a source of incoming partons (quarks
and gluons) and only a fraction of its momentum participates in the
hard scatter. In this case both quark and gluon
propagators are possible. 
\begin{figure}[h]
\begin{center}
\psfig{figure=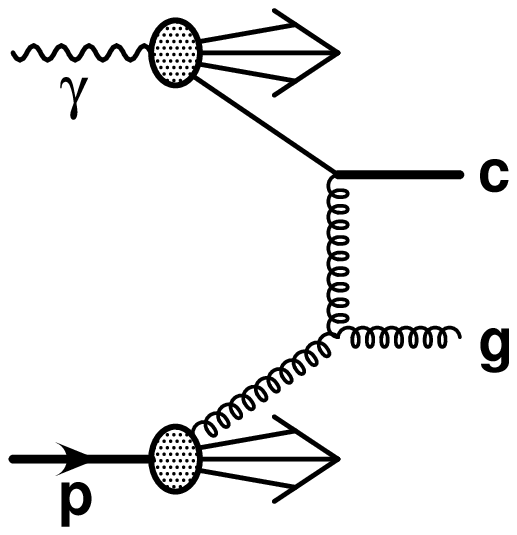,height=1.0in,width=1.2in}
\psfig{figure=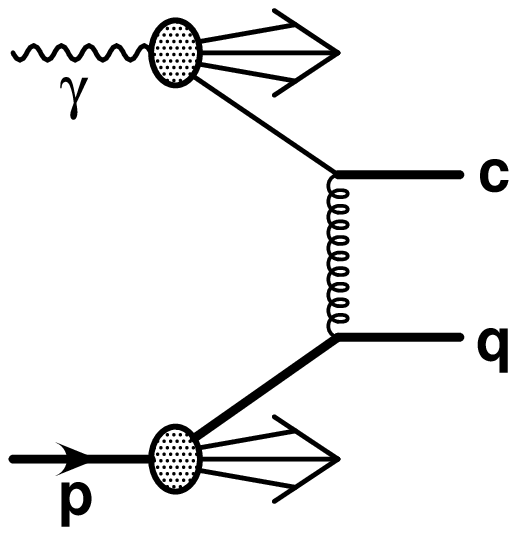,height=1.0in,width=1.2in}
\psfig{figure=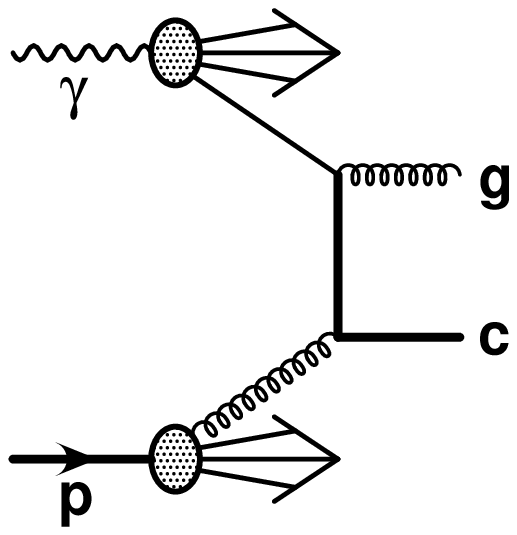,height=1.0in,width=1.2in}
\psfig{figure=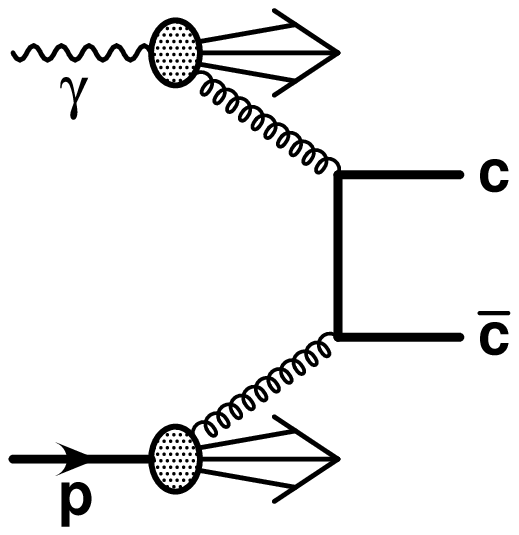,height=1.0in,width=1.2in}
\put(-310,-10){\small (a)}
\put(-220,-10){\small (b)}
\put(-130,-10){\small (c)}
\put(-40,-10){\small (d)}
\put(-310,-20){\line(0,0){6}}
\put(-310,-20){\line(1,0){194}}
\put(-210,-20){\line(0,1){6}}
\put(-115,-20){\line(0,1){6}}
\put(-260,-30){\small {\it Charm from photon}}
\end{center}
\vspace{-0.6cm}
\psfig{figure=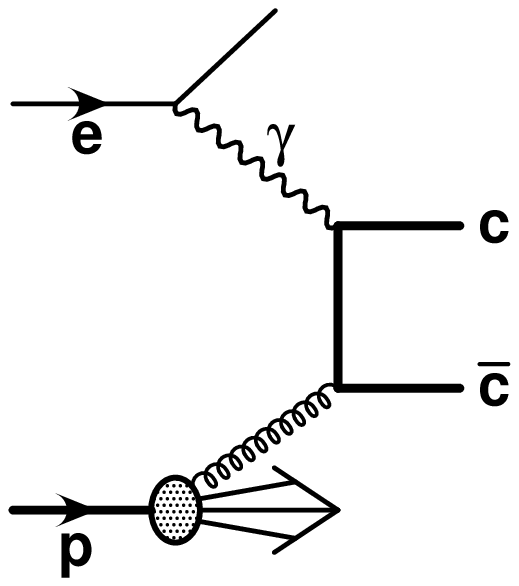,height=1.0in,width=1.2in}
\put(-40,-5){\small (e)}
\put(30,40){\small (a), (b), (c), (d) : {\it Resolved photon
    processes} }
\put(89,20){\small (e) : {\it Direct photon process} }

\caption {\it{Various sub-processes with charm, dominant in HERA region
    of phase space. \label{fig:1}}}
\end{figure}
\vspace{-0.6cm}
\begin{figure}[h]
\begin{center}
\psfig{figure=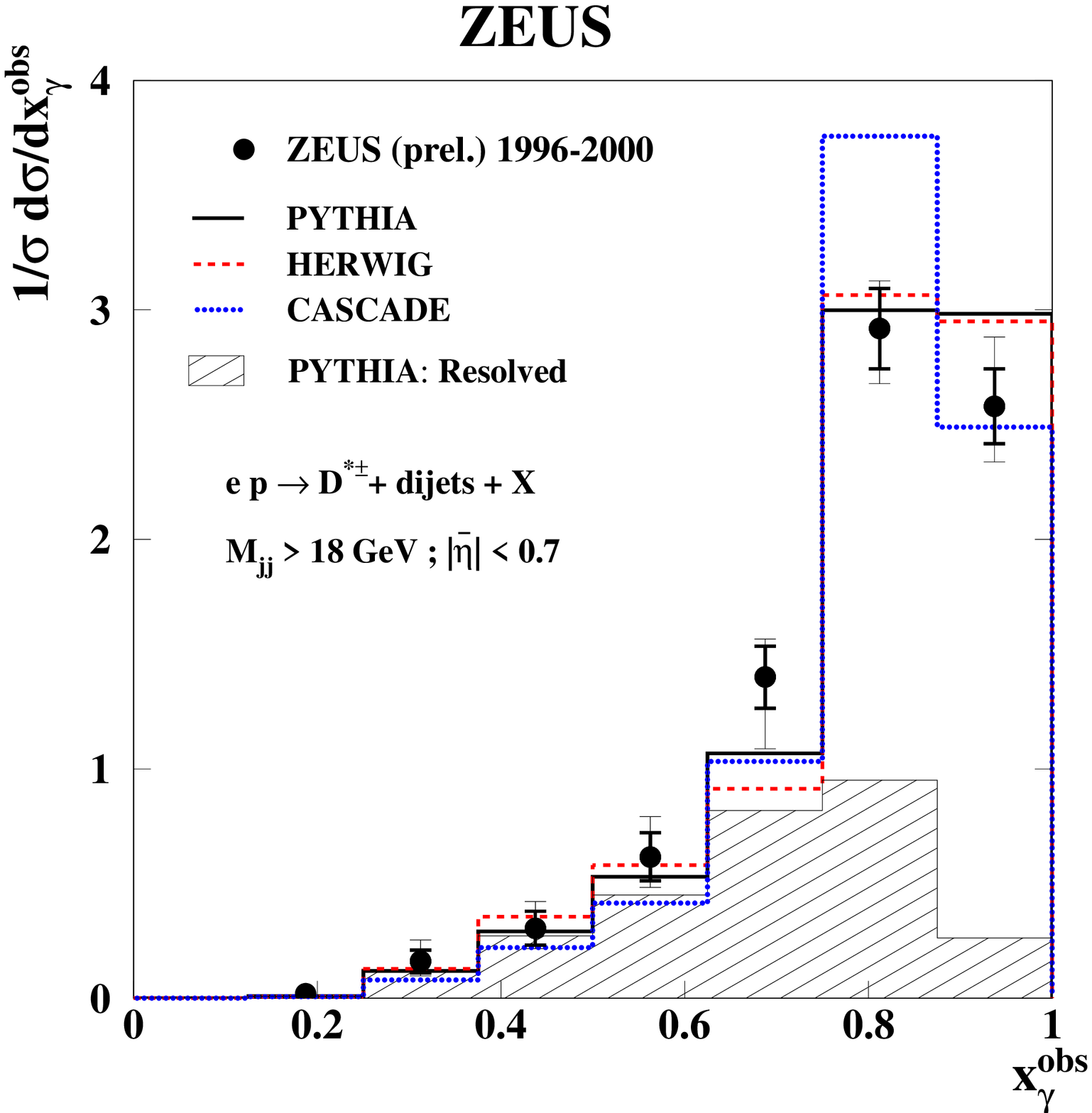,height=2.45in,width=2.45in}
\psfig{figure=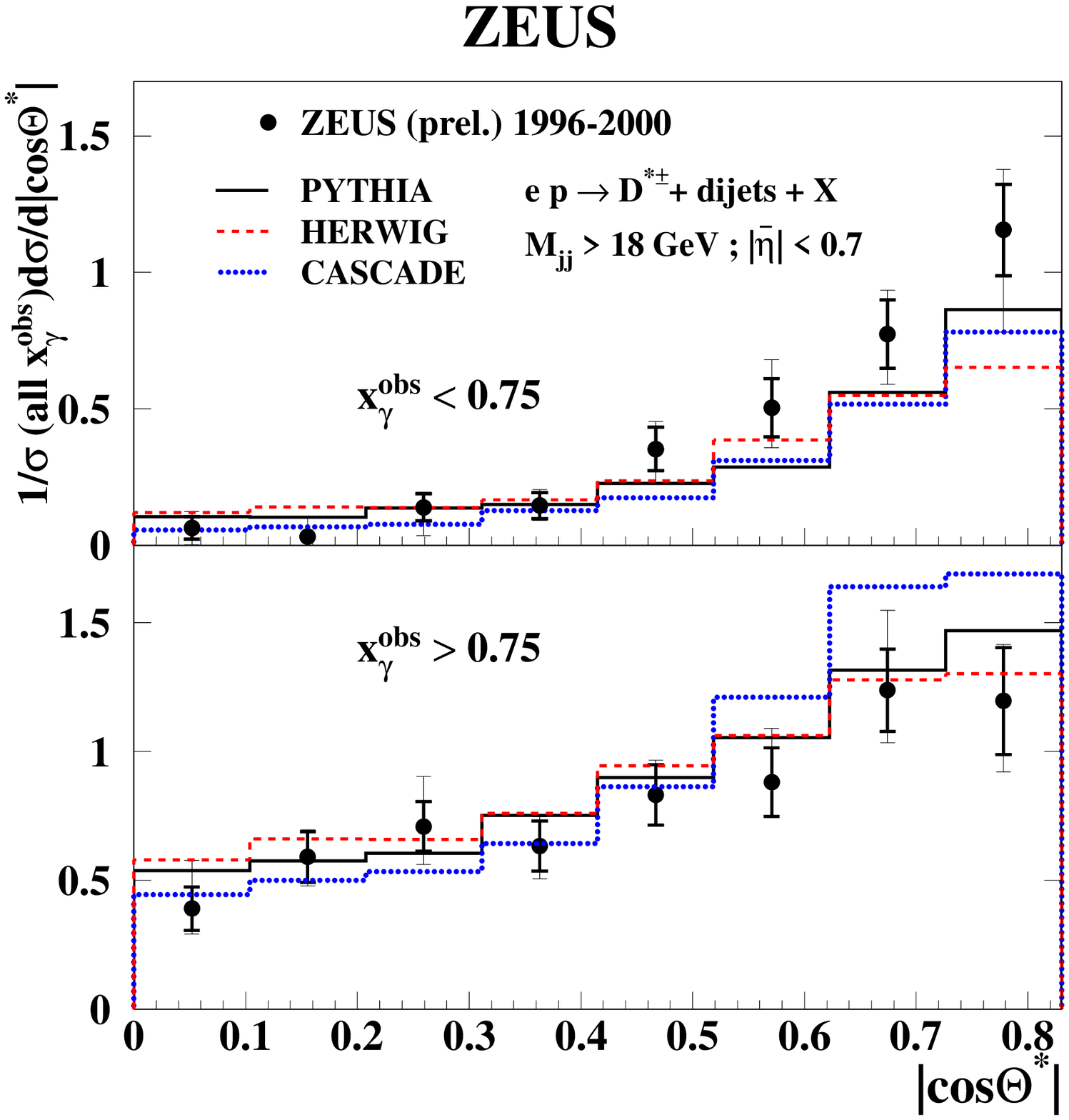,height=2.45in,width=2.45in}
\put(-320,-6){\small (a)}
\put(-140,-6){\small (b)}
\end{center}
\vspace{-0.5cm}
\caption {\it{Relative differential cross sections
    $1/\sigma d\sigma/dx_{\gamma}^{\rm obs}$(left), $1/\sigma$(all
    $x_{\gamma}^{\rm obs}$) $d\sigma/d|\cos\theta^{*}|$(right) with
    $p^{D*}_{\bot} > 3$ GeV, $|\eta^{D*}| < 1.5$, 130 $< W <$ 280 GeV,
    $|\eta^{\rm jet1,2}| < 2.4$, E$^{\rm jet1,2}_T > 5$ GeV, M$_{jj} >$ 18 GeV
    and $|\overline{\eta}| <$ 0.7. The results are compared with various
    MC simulations (histograms) }}
\end{figure}
In order to probe the charm dynamics in these
sub-processes and in particular to study the charm content of the
photon, the following measurements have been made.
\subsection{Measurement of $x_{\gamma}^{\rm obs}$}
The variable $x_{\gamma}^{\rm obs}$, related to the momentum fraction of
the parton from the photon, is defined as the fraction of the photon's
energy participating in the production of the two highest transverse
energy jets:
\begin{equation}
x^{\rm obs}_{\gamma} = \frac{\sum_{\rm jet1,2} E^{\rm jet}_{T}
  e^{-\eta^{\rm jet}}}{2yE_{e}}
\end{equation}
where $yE_{e}$ is the initial photon energy.
The normalised cross section as a function of
$x_{\gamma}^{\rm obs}$ is shown in Fig.2(a) compared with predictions
from PYTHIA~\cite{3}, HERWIG~\cite{4} and CASCADE~\cite{5}. As
previously mentioned, the photon acts
as a point like object for direct photon events, thus this
momentum fraction ($x_{\gamma}^{\rm obs}$) is expected to be populated
around high $x_{\gamma}^{\rm obs}$. The significant cross section at low 
$x_{\gamma}^{\rm obs}$ is consistent with the presence of resolved photon
processes. 

\subsection{Dijet angular distribution in $D^*$ photoproduction}

The high and low $x_{\gamma}^{\rm obs}$ region were studied in more
details in terms of dijet angular distribution, which are  
sensitive to the spin of the propagator. Fig.2(b) shows the relative
differential cross section as a funtion of $|\cos\theta^{*}|$, where $
\theta^{*}$ is the angle between the jet-jet axis 
and the beam direction in the dijet rest frame. The distributions are
enriched in direct photon ($x_{\gamma}^{\rm obs} > 0.75 $) or
resolved photon ($x_{\gamma}^{\rm obs} < 0.75 $) events. The measured
differential cross section $1/\sigma$(all $x_{\gamma}^{\rm obs}$)
$d\sigma/d|\cos\theta^{*}|$ for both of these samples are significantly
different,
indicating that the dominant mechanism for direct photon like events
proceed via $q-$exchange (spin$-1/2$ propagator, $\sigma \sim$ ($1 -
\cos\theta^{*}$)$^{-1}$), while resolved photon like events are
dominated by $g-$exchange (spin$-1$ propagator, $ \sigma \sim$ ($1 -
|\cos\theta^{*}|$)$^{-2}$, as in Rutherford scattering).\\

Most of the partonic processes with at least one charm in the final state, 
which contribute in the lowest order,
can be derived from Fig.1 by including other diagrams that are related
by crossing. Fig.3(a) shows the matrix element distribution for such
processes. As can be seen, the distributions that are symmetric in
$\cos\theta^*$ are $\gamma g \rightarrow c\overline{c} $ and $gg
\rightarrow c\overline{c}$; the other diagrams show asymmetry
either in the photon(negative) or proton(positive) hemisphere.
In order to study this behaviour of the matrix elements, the
charm-initiated jet, $D^*$ jet, and the other jet were separated in
$\eta - \phi$ space ($\triangle$R$_i$ 
$\equiv \sqrt{(\phi_{jet_{i}} - \phi_{D*})^{2} + (\eta_{jet_{i}} -
  \eta_{D*})^{2}}$ ; with $D^*$ jet having the smallest $\triangle$R$_{i(i=1,2)} < 1.0 $). Thus the sign of the unfolded
$\cos\theta^*$ distribution is given by the direction of the $D^*$
meson (positive for proton direction).

Fig.3(b) shows the differential
cross-section $1/\sigma$(all $x_{\gamma}^{\rm obs}$)
$d\sigma/d\cos\theta^{*}$ for both direct and resolved enriched
samples. The shaded areas for $x_{\gamma}^{\rm obs} < 0.75 $ and
$x_{\gamma}^{\rm obs} > 0.75 $ are, respectively, the contamination
of the genuine direct and resolved PYTHIA contributions.
The resolved enriched events in the photon hemisphere exhibit a strong
rise towards large negative $\cos\theta^*$ values, consistent with
$g-$exchange diagrams as expected 
from the matrix element distribution Fig.3(a). The only $g-$exchange
diagrams with this topology come from $ cg \rightarrow cg $ and $cq
\rightarrow cq$ (Fig.1(a)-(b)), where $c$ originates from the photon. On
the other hand, in the proton hemisphere, a mild rise towards large
positive $\cos\theta^*$ is 
observed, which is consistent with the $q-$exchange diagrams
(Fig.1(c)). These observations provide clear evidence that the bulk of the
resolved photon contribution is due to the charm
content of the photon, rather than to the more conventional resolved
process $gg \rightarrow c\overline{c}$.

The direct enriched events are consistent with $\gamma g \rightarrow
c\overline{c}$ (Fig.1(e)). The slight asymmetry in the $x_{\gamma}^{\rm
  obs} > 0.75 $ 
distribution can be accounted for by the contamination from genuine
resolved events, as given by PYTHIA.
\vspace{-0.3cm}
\begin{figure}[h]
\begin{center}
\psfig{figure=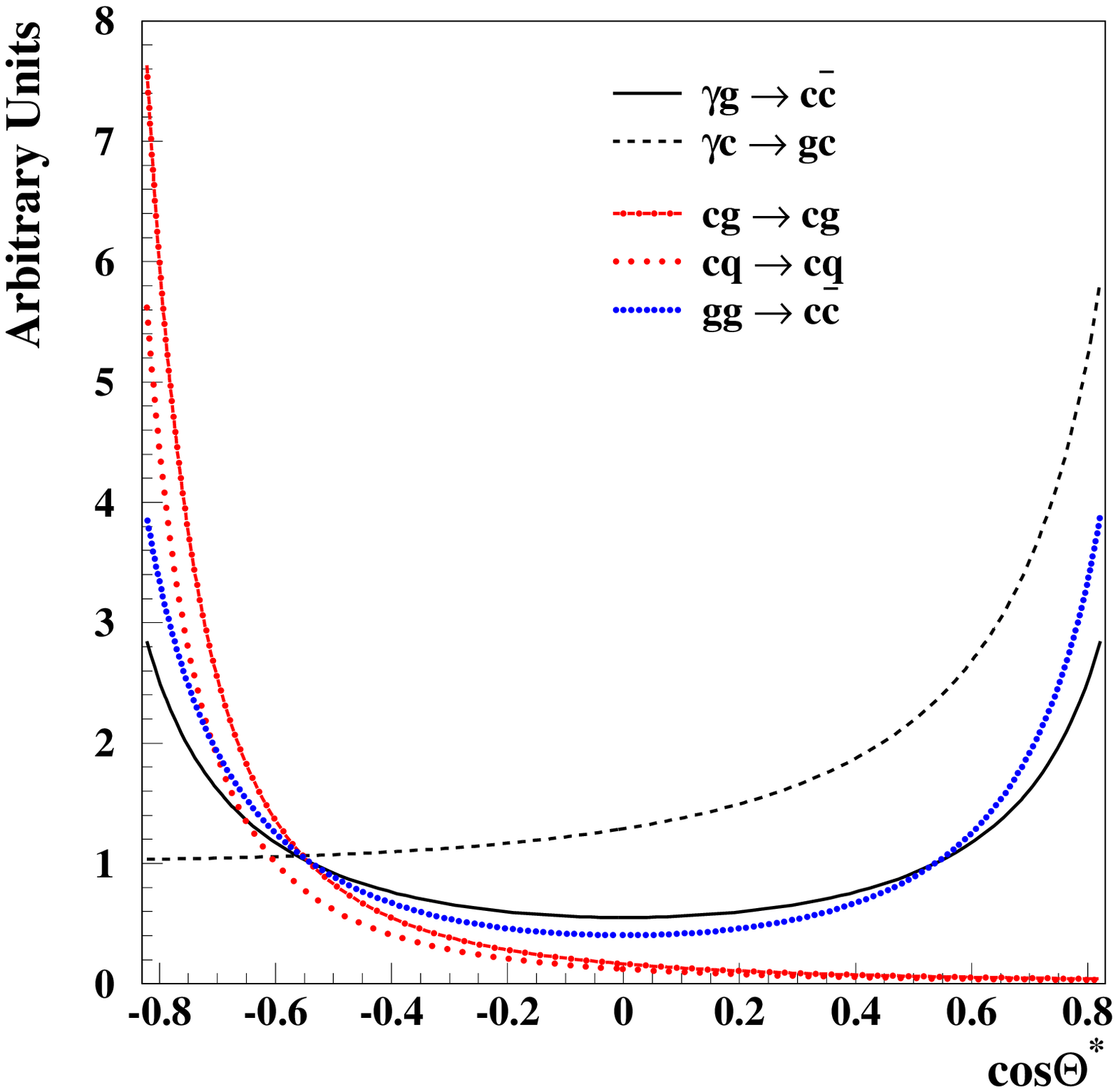,height=2.5in,width=2.2in}
\psfig{figure=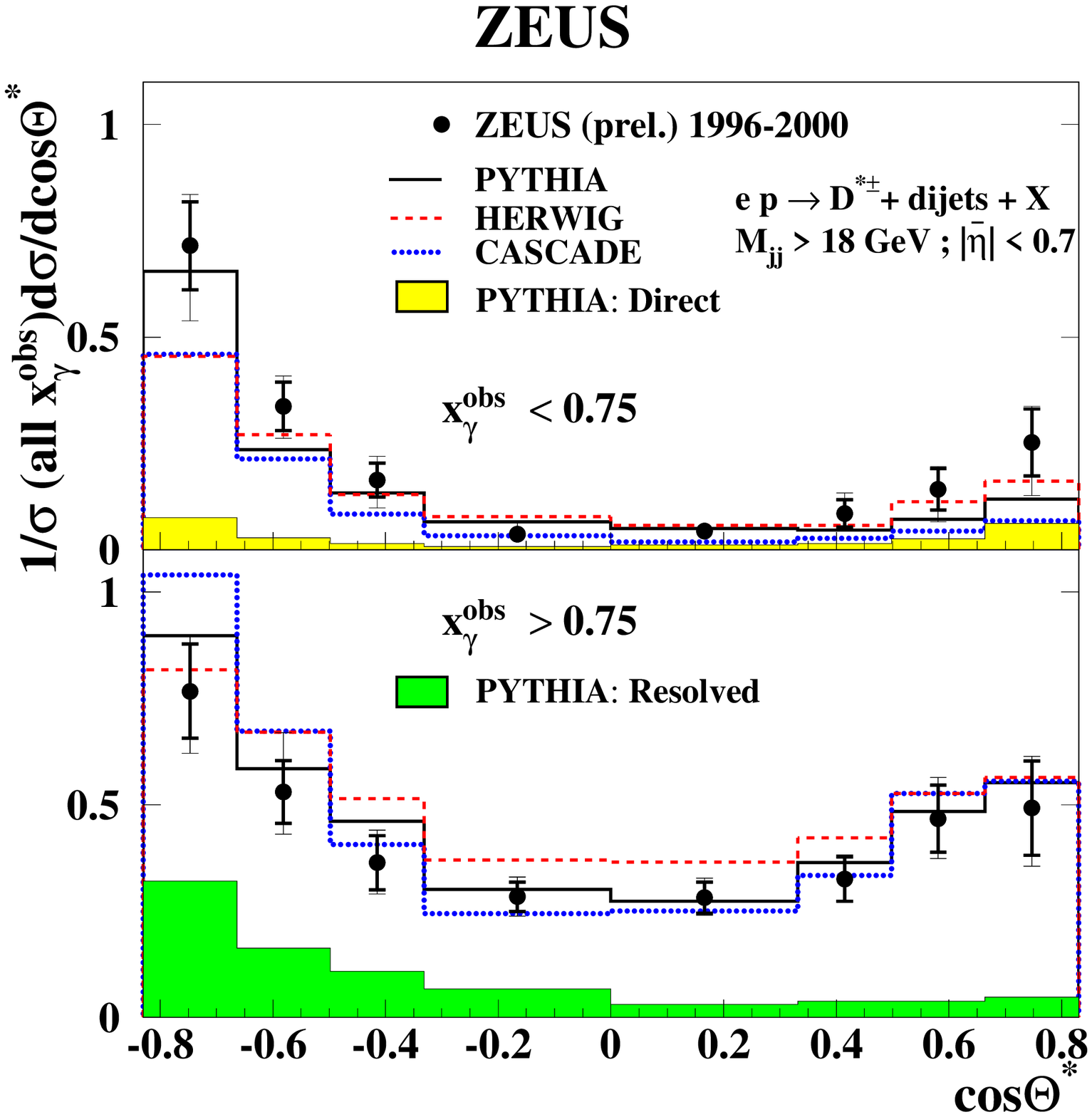,height=2.45in,width=2.7in}
\put(-310,-10){\small (a)}
\put(-150,-10){\small (b)}
\end{center}
\vspace{-0.3cm}
\caption {\it{ (a) The partonic cross section for $ 2 \rightarrow 2
    $ subprocesses, with at least a 
    charm in the final state, as a function of $\cos\theta^*$;
    (b) Relative differential cross sections $1/\sigma$(all 
    $x_{\gamma}^{\rm obs}$) $d\sigma/d\cos\theta^{*}$(right) in the same
    kinematic region as in Fig.2. The shaded area for $x_{\gamma}^{\rm
      obs} < 0.75 (x_{\gamma}^{\rm obs} > 0.75)$ plot is the
    contamination of the LO-direct (LO-resolved) photon contribution}}

\end{figure}

\section{Conclusions}
The measurement of P$_{v}$
is consistent with the universality of the charm fragmentation. 
Dijet angular distributions provide an
important tool for understanding the heavy quark production and
dynamics of underlying sub-processes. The
$\cos\theta^{*}$ distribution for dijet events with a $D^*$ shows a
clear signature of gluon propagator for events with $x^{\rm obs}_{\gamma}
< 0.75$, suggesting strong evidence that they mainly originate from
the charm content of the photon.

\end{document}